\documentclass[11pt]{article}
\usepackage[margin=1in]{geometry}
\usepackage{amsfonts,amsmath,amsthm,array,amssymb}
\usepackage{graphicx}
\usepackage{graphics}
\usepackage{multicol}
\usepackage{cite}
\usepackage{tikz}
\usepackage{relsize}
\usepackage[normalem]{ulem}
\usepackage{color}
\usepackage{epstopdf}
\usepackage{float}
\usepackage{nameref}
\usepackage{paralist}
\usepackage{caption}
\usepackage{subcaption}
\usepackage{hyperref}

\begin{document}


\title{Nowcasting Influenza Incidence with CDC Web Traffic Data: A Demonstration Using a Novel Data Set}
\author{Wendy K. Caldwell$^{1,2}$, Geoffrey Fairchild$^3$, Sara Y. Del Valle$^3$}
\date{}
\maketitle
\begin{center}
\footnotesize $^{1}$ School of Mathematical and Statistical Sciences, Arizona State University, Tempe, AZ\\
\vspace{.05in}
\footnotesize $^{2}$ X Computational Physics Division, Los Alamos National Laboratory, Los Alamos, NM\\
\vspace{.05in}
\footnotesize $^{3}$ Analytics, Intelligence, and Technology Division, Los Alamos National Laboratory, Los Alamos, NM
\end{center}
\vspace{.25in}
\hrule
\vspace{.25in}

\begin{abstract}
Influenza epidemics result in a public health and economic burden around the globe. Traditional surveillance techniques, which rely on doctor visits, provide data with a delay of 1--2 weeks. A means of obtaining real-time data and forecasting future outbreaks is desirable to provide more timely responses to influenza epidemics. {In this work, we present the first implementation of a novel data set by demonstrating its ability to supplement traditional disease surveillance at multiple spatial resolutions.} We use Internet traffic data from the Centers for Disease Control and Prevention (CDC) website to determine the potential usability of this data source. We test the traffic generated by ten influenza-related pages in eight states and nine census divisions within the United States and compare it against clinical surveillance data. Our results yield $r^2$ = 0.955 in the most successful case, promising results for some cases, and unsuccessful results for other cases. These results demonstrate that Internet data may be able to complement traditional influenza surveillance in some cases but not in others. Specifically, our results show that the CDC website traffic may inform national and division-level models but not models for each individual state. In addition, our results show better agreement when the data were broken up by seasons instead of aggregated over several years. In the interest of scientific transparency to further the understanding of when Internet data streams are an appropriate supplemental data source, we also include negative results (i.e., unsuccessful models). We anticipate that this work will lead to more complex nowcasting and forecasting models using this data stream.
\end{abstract}

\section{Introduction}
Every year, an estimated 5\% to 20\% of people in the United States become infected with influenza \cite{hickmann2015forecasting}. The typical influenza season begins in October and ends in May, with the peak occuring in the winter months. Annually, 3,000--50,000 people die from the flu, with another 200,000 requiring hospitalization {\cite{molinari2007annual}}. The yearly flu burden is estimated to cost around \$87 billion in lost productivity {\cite{molinari2007annual}}. Timely surveillance of influenza can help reduce this burden, allowing health care facilities to more adequately prepare for the influx of patients when flu levels are high \cite{dugas2013influenza}.\\
\indent One common surveillance measure is the fraction of patients presenting with influenza-like illness (ILI), consisting of a fever of at least 100$^{\text{o}}$ F ($37.8^{\text{o}}$ C) and a cough or sore throat with no other known cause {\cite{fluglossary}}. ILI data are collected from about 2,900 volunteer health care providers throughout the United States, although each week only about 1,800 report their data. These data are then aggregated and made public, after a time lag of about 1--2 weeks \cite{reid, nick, mciver2014wikipedia, hickmann2015forecasting, polgreen2008using, kim2013use}. {Because the ILI data are collected from volunteer providers, the data set is incomplete. If policies were enacted to provide incentive for reporting health care providers, or to make reporting compulsory, the result would be a more complete data set. Other surveillance systems include virological data from the World Health Organization, emergency department visits, electronic health records, crowd-sourced ILI reports, Widely Internet Sourced Distributed Monitoring, Influenzanet, and Flu Near You  \cite{baltrusaitis2018comparison,brownstein2017combining}.}

	\subsection{Internet data streams}
	In the United States, 87\% \cite{pewhfs} of adults use the Internet. Of those Internet users, 72\% \cite{pewhfs} have used the Internet to search for health information within the last year. The most common health-related searches are for information regarding a specific disease or condition (66\%) and information about a specific treatment or procedure (56\%) \cite{pewhfs, pewht}. \\
\indent There are two main types of health-related Internet activity. The first is health sharing, in which Internet users post about health-related topics (e.g., a tweet about being sick). The second is health seeking, in which users utilize the Internet to obtain information about health-related topics \cite{reid}. In this paper, we focus on health-seeking behavior. Previous studies have shown that analyzing online health-seeking behavior can improve early detection of disease incidence by detecting changes in disease activity {\cite{ginsberg2009detecting, polgreen2008using, chretien2014influenza, perrotta2017using, shaman2012forecasting, lampos2017enhancing}}. Similarly, other studies have shown that Internet data emerging from search queries can aid detection of outbreaks in areas with large populations of Internet users \cite{xu2011predicting}, because online health-related search queries and epidemics are often strongly correlated \cite{xu2011predicting, jia2013gonorrhea}.\\
\indent Internet data have been used to forecast disease incidence in other models. Polgreen et al. developed {linear} influenza forecasting models with lags of 1 to 10 weeks for each of the 9 U.S. census regions {using search queries from Yahoo} \cite{polgreen2008using}. The best performing models had lags of 1--3 weeks and an average $r^2$ of 0.38 (with a high of 0.57 in the East-South-Central region) \cite{polgreen2008using}. These low $r^2$ values demonstrate potential problems in relying on search information alone. Ginsberg et al. were able to predict influenza epidemics two weeks in advance {using Google search queries to fit linear models using log-odds of ILI visits and related searches} \cite{ginsberg2009detecting}. \\
\indent Using a Poisson distribution and LASSO regression, McIver and Brownstein obtained an $r^2$ value of 0.946 using Wikipedia data \cite{mciver2014wikipedia}, although some data were excluded from analyses due to increased media attention and higher than normal influenza activity. Generous et al. used Wikipedia data to train a statistical model with linear regression, which {demonstrated its potential for} forecasting disease incidence around the globe, including influenza in the United States, which had an $r^2$ of 0.89 \cite{nick}. Hickmann et al. conducted a similar study {of linear regression models} which showed that using Wikipedia to forecast influenza in the United States for the 2013--2014 season resulted in an $r^2$ value greater than 0.9 in some instances \cite{hickmann2015forecasting}. \\
\indent Integrating both Wikipedia data and Google Flu Trends, Bardak et al. obtained $r^2$ values of 0.94 and 0.91 using ordinary least squares (OLS) and ridge regression, respectively, for forecasting influenza outbreaks \cite{bardak2015prediction}. For OLS nowcasting, the $r^2$ value was 0.98 in the best case. For the best fit, the weekly data was offset by one week \cite{bardak2015prediction}.\\
\indent As part of the CDC's 2013--2014 Predict the Influenza Season Challenge, 9 teams used digital data sources to create forecasting models. The digital sources these teams utilized were Wikipedia, Twitter, Google Flu Trends, and HealthMap. The teams used either mechanistic or statistical models to create their forecasts, with the most successful team using multiple data sources, which may have reduced biases usually associated with Internet data streams \cite{biggerstaff2016results}. Broniatowski et al. used Twitter data to detect increasing and decreasing influenza prevalence with 85\% accuracy \cite{broniatowski2013national}. Zhang et al. used Twitter data to inform stochastically, spatially structured mechanistic models of influenza in the United States, Italy, and Spain \cite{zhang2017forecasting}.\\
\indent Internet data streams have also been used to supplement traditional surveillance techniques with nowcasting models. Paul et al. used Twitter along with ILI data from the CDC to produce nowcasting influenza models as well as nowcasting models using solely ILI data. They conclude that the addition of Twitter data led to more accurate nowcasting models \cite{paul2014twitter}. Santillana et al. combined Google Trends data and CDC-reported ILI data to create models for nowcasting and forecasting influenza \cite{santillana2015combining}. Lampos et al. used search query data to explore both linear and nonlinear nowcasting models \cite{lampos2015advances}. Yang et al. used Google search data to create an influenza tracking model with autoregression \cite{yang2015accurate}. \\
\indent In contrast, we consider data on page views of the CDC website rather than search data from sites not solely devoted to public health. {We use this data set because we expect it to be inherently less noisy because of its focus on public health issues.} We use ordinary least squares to nowcast influenza nationally, across the 9 U.S. census divisions, and across 8 states using access data from 10 influenza-related CDC pages. {Our nowcasting models cover influenza seasons from 2013 to 2016, with the 2012--2013 season being partially included because our data set begins Jan. 1, 2013. The inclusion of an incomplete influenza season serves to inform whether this data set can be used given a more restrictive time frame.} We include both positive and negative results to advance our knowledge regarding when Internet data may or may not work. The negative results are crucial to advancing the field of disease surveillance using Internet data, as they demonstrate when these data sources contribute to unreliable surveillance. We focus on answering the following two research questions:\\

\textbf{Q1}: Can CDC page visits be used as an additional data source for monitoring disease incidence?\\

\vspace{.25cm}

\textbf{Q2}: What is the appropriate shift needed to obtain the best data fit?

\section{Methods}

	\subsection{Data Sources}
	We used page view data provided by the Centers for Disease Control and Prevention (CDC). Each data point contains the page name, date and time of access, and the geographic location from where the page was viewed. {These data are available at geographic resolutions of national and state levels and include some metropolitan areas (e.g., New York City). The data are available at a number of temporal resolutions beginning on January 1, 2013. For these models, we use weekly page view data to coincide with the ILI data temporal resolution. The data are available as raw page view counts and normalized page view counts, and we consider the latter for this work.} We selected pages associated with general influenza information, treatment, and diagnosis. Pages were sometimes renamed, but we were able to follow the evolution of each selected page by utilizing key words in the page titles as well as the date ranges for available data.\\
\indent Because the majority of health-related Internet searches concern specific conditions, treatments, and procedures \cite{pewht}, we selected pages related to those topics. These pages also align with Johnson et al., who used pages in the categories of Diagnosis/Treatment and Prevention/Vaccination for influenza surveillance \cite{johnson2004analysis}. Specifically, we used the following pages: antivirals, flu basics, FluView, high risk complications, key facts, prevention, symptoms, treating influenza, treatment, and vaccine. {We then aggregate the page views of interest for each of our models.}  A complete list of pages can be found in \nameref{Appendix A}.\\
\indent The states we selected were based on severity of flu {(determined from FluView)} during the available seasons and the availability of ILI data{{, which is not standardized and is dependent on each state's reporting mechanism}. ILI data for each state include the week ending or starting date as well as the percentage of influenza-like illness for the specified week. While some states also report additional data, such as school closures and hospitalizations, these data are not made available by every state. Note that the ILI reporting and accessibility vary across all the states. The states we selected were \begin{inparaenum}[1)] \item California, \item Maine, \item Missouri, \item New Jersey, \item New Mexico, \item North Carolina, \item Texas, and \item Wisconsin\end{inparaenum}. With the exception of Texas, these states did not release ILI data outside of the typical flu season. A complete list of the data sources for the state ILI can be found in \nameref{Appendix B}, and the clinical data are available in \nameref{Appendix E}.\\
\indent Fig \ref{iliplot} shows the percentage of ILI visits for each state considered in this study as well as the national percentage of ILI visits. We see distinct spikes that {indicate} the peaks of the flu seasons. With the exception of Maine, which behaves as an outlier at times, the figure shows spikes indicating there are ``peak" weeks for influenza-related page views. {Texas also exhibits outlier behavior with ILI percentages consistently higher than the typical national baseline of 2\%, which is used to determine when the flu has reached epidemic status. These two outliers are shown in teal (Texas) and dark blue (Maine). The national ILI is shown in black. The remaining states exhibit behavior consistent with the national ILI trend.} Fig \ref{cdcdens} shows the CDC page view data as a heat map: weeks with more page views are shown darker than weeks with fewer page views.
\begin{figure}[H]
\begin{center}
\includegraphics[width=.75\textwidth]{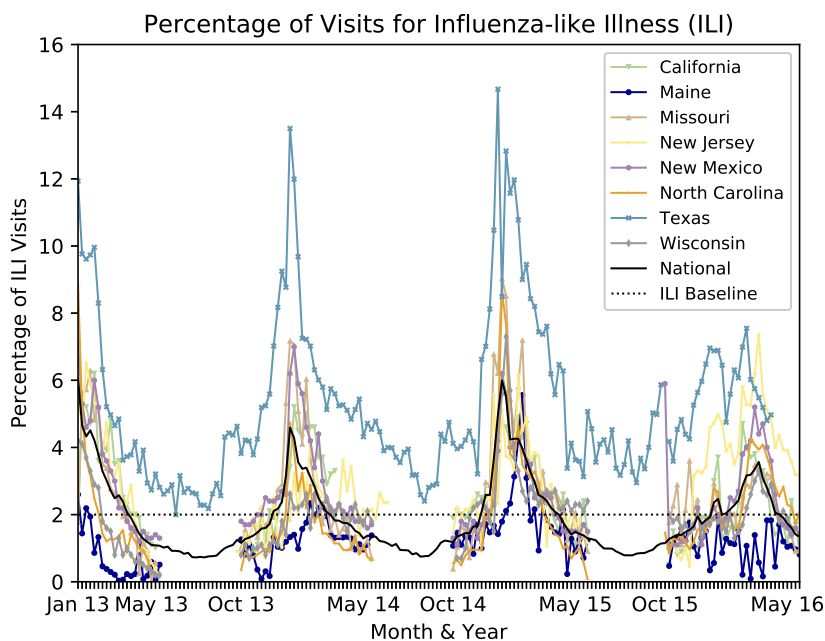}
\end{center}
\caption{Percentage of ILI visits per state, compared to the typical national baseline of 2\%. Maine {(dark blue)} and Texas {(teal)} exhibit outlier behavior, with Texas having a greater ILI percent and Maine having a lesser ILI percent. {The remaining states follow the national ILI trend, shown in black.}}
\label{iliplot}
\end{figure}
\begin{figure}[H]
\begin{center}
\includegraphics[width=.75\textwidth]{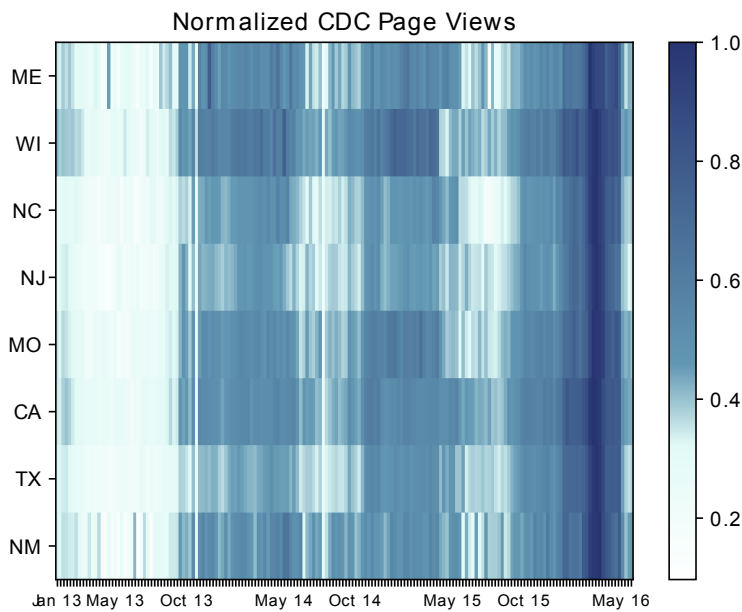}
\end{center}
\caption{This figure shows normalized CDC web traffic as a heat map. Darker areas indicate more page views and appear to correlate with increases in ILI. The page views also appear to be more prevalent during the typical influenza season, October--May.}
\label{cdcdens}
\end{figure}

\noindent
In addition to selected states, we also considered the 9 U.S. census divisions: New England, Middle Atlantic, East North Central, West North Central, South Atlantic, East South Central, West South Central, Mountain, Pacific. For a list of states included in each division, see \nameref{Appendix C}. Data for the census divisions {were} obtained from the CDC and can be viewed in \nameref{Appendix D}.

	\subsection{Linear Regression}
	We used statsmodels version 0.9.0, a statistical analysis module for Python, to perform linear regression on our data sets using OLS. This creates a linear model {$M$} of the form
	\begin{equation}
	M = \sum_{i=0}^n \alpha_iX_i, \nonumber
	\end{equation}
where $\alpha_i$ are the regression coefficients, and $X = \left(1, X_1, X_2, ..., X_n\right)$ is the vector of CDC page view data, with $n$ representing the number of CDC pages used for the model, ranging from 1 to 10. We correlate ILI and CDC page views for the same week or with a one-week shift. In the shifted cases, we shift the ILI data forward by one week, so that the model associates the current week's page views with the following week's ILI data. This shifting is performed to account for the incubation period of influenza and the time between the onset of symptoms and the first doctor visit. Statsmodels uses the CDC page view and ILI data to determine the appropriate regression coefficients, fits parameters with OLS, and computes the goodness of fit, $r^2$, also referred to as the coefficient of determination. The $r^2$ value measures how well two time series correlate. An $r^2 = 1$ indicates a perfect fit, while an $r^2$ value of 0 indicates no correlation. Although $r^2$ is not necessarily the best metric to use for judging goodness of fit \cite{reid}, it is nonetheless the most common metric used and still provides one with a decent overall sense of fit quality. Additionally, we examined the root mean squared error (RMSE) and the normalized root mean squared error (NRMSE) using Python's sklearn libraries.

\section{Results}
We analyzed the data at the national, division, and state levels and computed the $r^2$ for each geographic resolution. In this section, we discuss the results of our experiments, both successes and failures. We include figures of models at the national, census division, and state levels. Because of the varying scales between page views and ILI percent, we choose to normalize the data and our models in order to plot them on the same axes. We use raw data to create the models, and then we normalize the each model with respect to its maximum. We also normalize the ILI data and CDC.gov web traffic data with respect to their maximums for the given time period so that all three curves may appear in the same plot.

	\subsection{National Results}
	We selected pages that corresponded to the topics most often searched during online health-seeking activites. When we combined all ten pages, we were able to achieve an $r^2$ value of 0.889 for the national 2012--2013 influenza season after implementing a one-week shift. We also had success modeling the national 2015--2016 influenza season with no shift, achieving an $r^2$ value of 0.834. We obtained better results when limiting the pages to FluView, Symptoms, and Treatment{, which we attribute to the information on these pages aligning with topics most commonly used for Internet health seeking}. For these pages, the most successful models did not have a shift. For the 2012--2013 influenza season, we achieved an $r^2$ of 0.906. The model for the 2015--2016 season had an $r^2$ value of 0.891. Table \ref{natltable} shows the most successful model for each influenza season included in this study. Fig. \ref{nat} shows these models.
\begin{table}[H]
\centering
\begin{tabular}{|c|c|c|c|c|c|}
\hline
\textbf{Pages} & \textbf{Season} & \textbf{Shift} & $\mathbf{r^2}$ & {RMSE} & {NRMSE}\\
\hline
FluView, Symptoms, Treatment & 2012--2013 & None & 0.912 & {0.423} & {0.070}\\
\hline
Symptoms & 2015--2016 & None & 0.892 & {0.213} & {0.060}\\
\hline
FluView & 2013--2014 & None & 0.802 & {0.510} & {0.111} \\
\hline
Antivirals, Prevention & 2014--2015 & None & 0.778 & {0.615} & {0.103}\\
\hline
\end{tabular}
\caption{This table lists the pages and shift for the most successful models for each influenza season at the national level.}
\label{natltable}
\end{table}
\begin{figure}[H]
\centering
\begin{subfigure}[b]{0.49\textwidth}
\includegraphics[width=\textwidth]{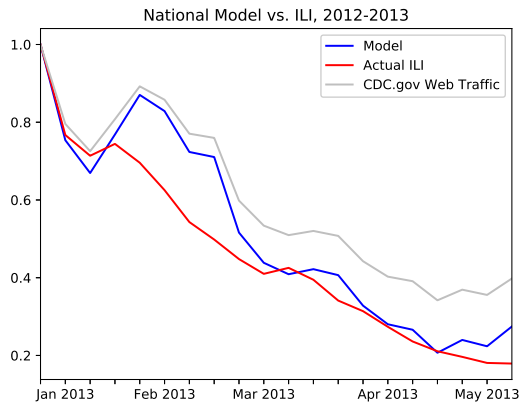}
\caption{FluView, Symptoms, Treatment, 2012--2013}
\label{natfvst}
\end{subfigure}
\begin{subfigure}[b]{0.49\textwidth}
\includegraphics[width=\textwidth]{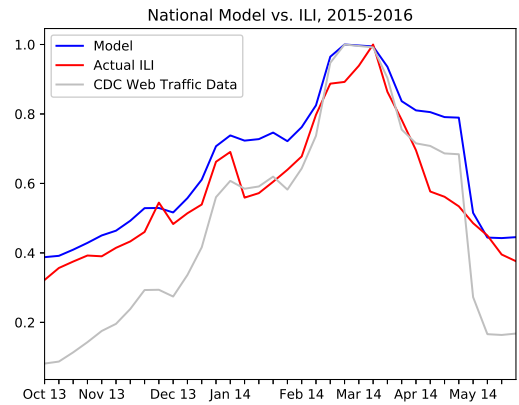}
\caption{Symptoms, 2015--2016}
\label{nats}
\end{subfigure}
\begin{subfigure}[b]{0.49\textwidth}
\includegraphics[width=\textwidth]{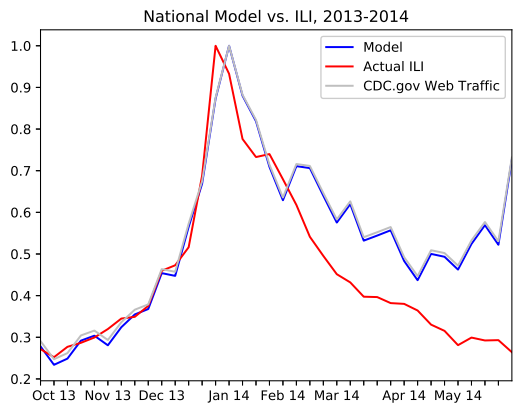}
\caption{FluView, 2013--2014}
\label{natfv}
\end{subfigure}
\begin{subfigure}[b]{0.49\textwidth}
\includegraphics[width=\textwidth]{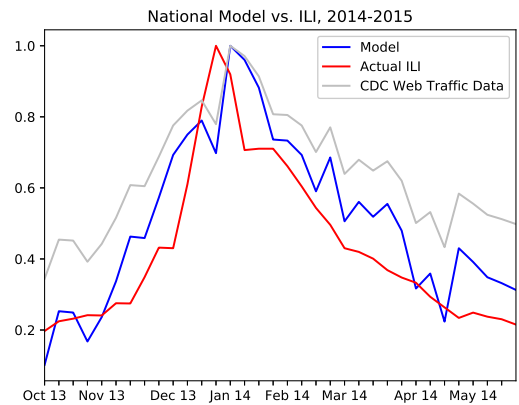}
\caption{Antivirals, Prevention, 2014--2015}
\label{natap}
\end{subfigure}
\caption{These plots show national models and the associated pages and influenza seasons.}
\label{nat}
\end{figure}
	
	\subsection{Census Division Results}
	Using the data for each of the nine census divisions, we were able to achieve $r^2 > 0.7$ in at least one case for each division. We considered all seasons together and separately, with the better results coming from modeling each individual season. We considered all pages together and pages most closely associated with topics most commonly searched by health-seeking individuals. In the most successful case, the model was able to closely match the 2015--2016 influenza season for the West North Central division with an $r^2$ of 0.955 using the FluView, Symptoms, and Treatment pages. Although we had successes using all 10 pages, the most successful model for each division involved only these three pages. Fig. \ref{reg} shows some of these models, and Table \ref{divother} highlights these successes.
\begin{figure}[H]
\centering
\begin{subfigure}[b]{0.49\textwidth}
\includegraphics[width=\textwidth]{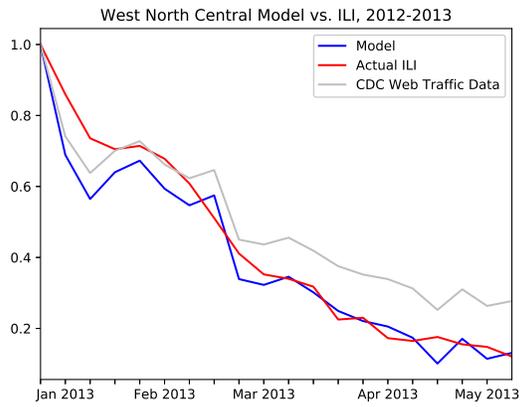}
\caption{West North Central, 2012--2013}
\end{subfigure}
\begin{subfigure}[b]{0.49\textwidth}
\includegraphics[width=\textwidth]{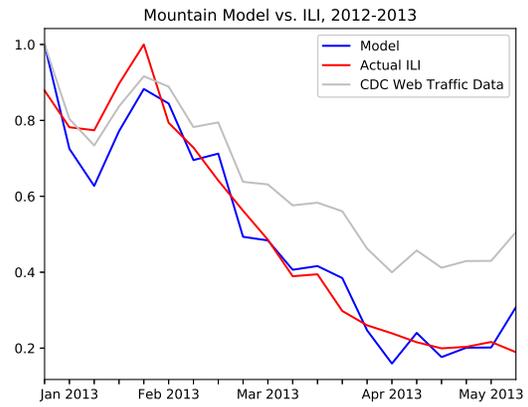}
\caption{Mountain, 2012--2013}
\end{subfigure}
\begin{subfigure}[b]{0.49\textwidth}
\includegraphics[width=\textwidth]{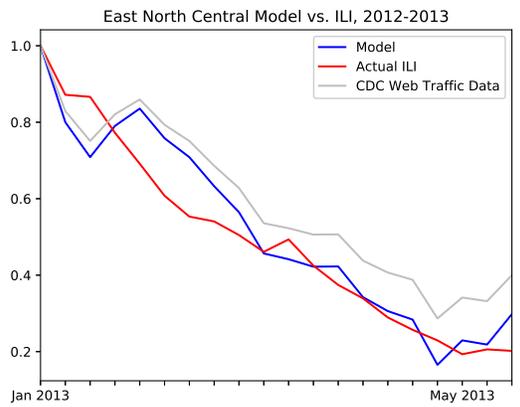}
\caption{East North Central, 1 Week Shift, 2012--2013}
\end{subfigure}
\begin{subfigure}[b]{0.49\textwidth}
\includegraphics[width=\textwidth]{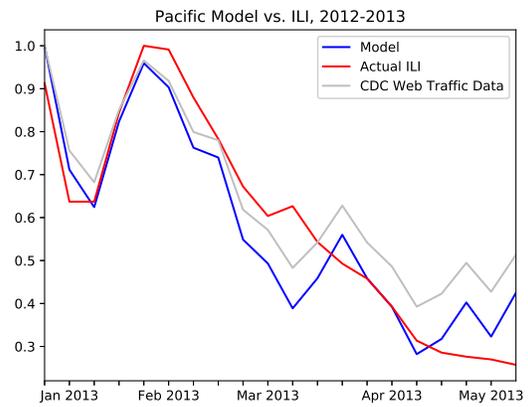}
\caption{Pacific, 2012--2013}
\end{subfigure}
\begin{subfigure}[b]{0.49\textwidth}
\includegraphics[width=\textwidth]{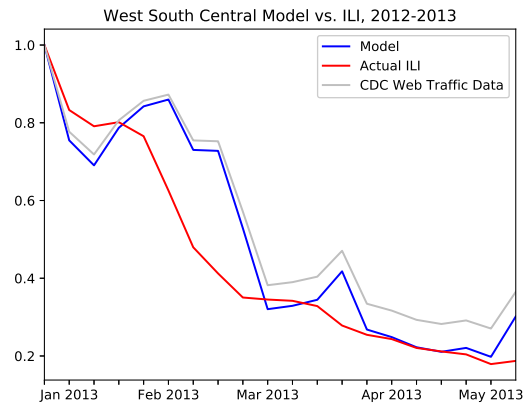}
\caption{West South Central, 2012--2013}
\end{subfigure}
\caption{These plots show census division model successes using the FluView, Symptoms, and Treatment pages for the 2012--2013 influenza season.}
\label{reg}
\end{figure}
\begin{table}[H]
\centering
\begin{tabular}{|c|c|c|c|c|c|}
\hline
\textbf{Division} & \textbf{Season} & \textbf{Shift} & $\mathbf{r^2}$ & RMSE & NRMSE\\
\hline
West North Central & 2012--2013 & None & 0.955 & 0.367 & 0.057\\
\hline
Mountain & 2012--2013 & None & 0.921 & 0.336 & 0.077\\
\hline
New England & 2015--2016 & None & 0.920 & 0.096 & 0.096\\
\hline
East North Central & 2012--2013 & 1 Week & 0.899 & 0.331 & 0.076\\
\hline
South Atlantic & 2015--2016 & None & 0.893 & 0.218 & 0.065\\
\hline
Middle Atlantic & 2015--2016 & None & 0.861 & 0.302 & 0.073\\
\hline
Pacific & 2012--2013 & None & 0.849 & 0.503 & 0.094\\
\hline
West South Central & 2012--2013 & None & 0.828 & 0.986 & 0.105\\
\hline
East South Central & 2015--2016 & 1 Week & 0.793 & 0.365 & 0.082\\
\hline
\end{tabular}
\caption{This table shows each of the 9 census divisions and the season and shift for which the division's model had the highest $r^2$ value. The table also shows the root mean squared error (RMSE) and the normalized root mean squared error (NRMSE). The results presented correspond to the FluView, Symptoms, and Treatment pages aggregated.}
\label{divother}
\end{table}

	\subsection{State Results}
	We found $r^2$ for each of the states considered in this study, using a variety of pages and page combinations. Table \ref{statetable} lists the most successful model for each state, the season, the data shift, and the $r^2$ value.
\begin{table}[H]
\centering
\begin{tabular}{|c|c|c|c|c|c|c|}
\hline
\textbf{State} & \textbf{Page(s)} & \textbf{Season} & \textbf{Shift} & $\mathbf{r^2}$ & {RMSE} & {NRMSE}\\
\hline
Texas & All & $2012-2013$ & 1 week & {0.917} & {0.730} & {0.073}\\
\hline
Wisconsin & FVST & $2012-2013$ & None & 0.833 & {0.533} & {0.127} \\
\hline
New Jersey & All & $2012-2013$ & 1 week & 0.832 &{0.767} &{0.117} \\
\hline
Missouri & FVST & $2012-2013$ & 1 week & 0.823 &{0.801}&{0.127} \\
\hline
North Carolina & FVST & $2015-2016$ & 1 week & 0.781 &{0.455} &{0.106} \\
\hline
New Mexico & All & $2015-2016$ & 1 week & 0.771 &{1.184} & {0.197} \\
\hline
California & FVST & $2012-2013$ & 1 week & 0.758 & {0.777}& {0.125}\\
\hline
Maine & Antivirals & $2012-2013$ & None & 0.662 & {0.445}& {0.171}\\
\hline
\end{tabular}
\caption{This table shows the most successful results for each state considered in this study. ``All" refers to an aggregation of all 10 pages, and ``FVST" refers to an aggregation of the FluView, Symptoms, and Treatment pages.}
\label{statetable}
\end{table}

Fig \ref{states} shows both successes and failures at the state level. Adding all of the pages together, we were able to obtain $r^2$ values of {0.917} and 0.801 for Texas (see Fig \ref{alltx}) and Wisconsin {(see Fig \ref{allwiseas1})}, respectively, for the 2012--2013 influenza season. For the 2013--2014 season, the highest $r^2$ value was 0.187 for Wisconsin (see Fig \ref{allwi}). For the 2014--2015 season, the highest $r^2$ value was 0.322 for Missouri (see Fig \ref{allmo}). For the 2015--2016 season, the highest $r^2$ value was 0.647 for North Carolina (see Fig \ref{allnc}). 
\begin{figure}[H]
\centering
\begin{subfigure}[b]{0.49\textwidth}
\includegraphics[width=\textwidth]{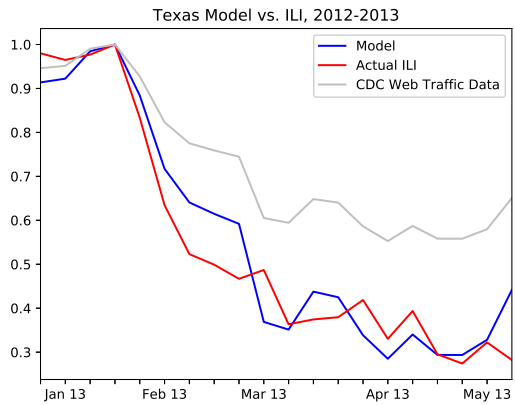}
\caption{Texas, 1 Week Shift, 2012--2013}
\label{alltx}
\end{subfigure}
\begin{subfigure}[b]{0.49\textwidth}
\includegraphics[width=\textwidth]{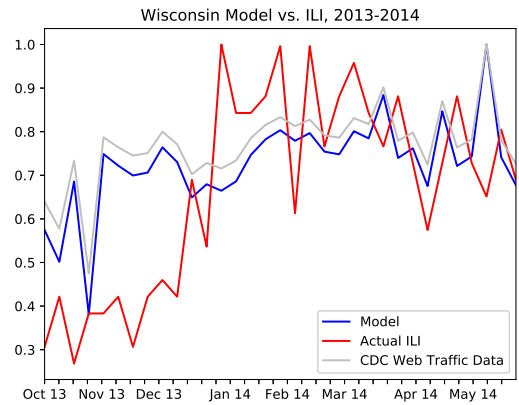}
\caption{Wisconsin, 2013--2014}
\label{allwi}
\end{subfigure}
\begin{subfigure}[b]{0.49\textwidth}
\includegraphics[width=\textwidth]{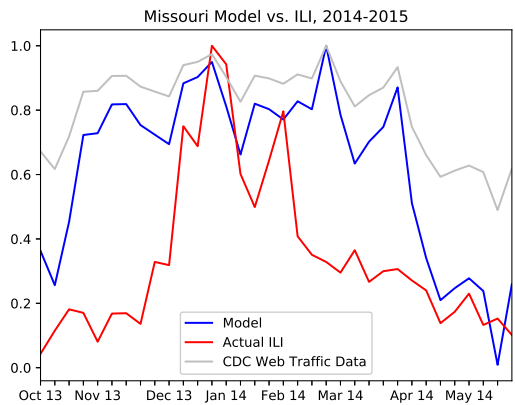}
\caption{Missouri, 2014--2015}
\label{allmo}
\end{subfigure}
\begin{subfigure}[b]{0.49\textwidth}
\includegraphics[width=\textwidth]{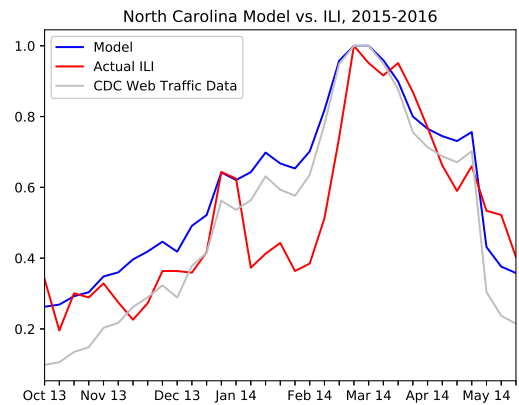}
\caption{North Carolina, 2015--2016}
\label{allnc}
\end{subfigure}
\begin{subfigure}[b]{0.49\textwidth}
\includegraphics[width=\textwidth]{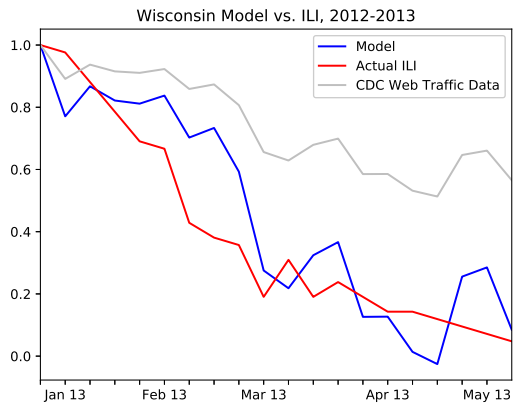}
\caption{Wisconsin, 2012--2013}
\label{allwiseas1}
\end{subfigure}
\caption{These plots show different states during different seasons. The $r^2$ values of each of these models ranges from 0.187 to {0.917}. These models aggregated all 10 pages, and the success varied by state.}
\label{states}
\end{figure}

We were not surprised that Texas had the best fit. Texas was the only state we included that provided ILI data not only for the typical influenza season but also for the off-season. This additional data likely contributed to the success of the Texas models. {The lack of success we encountered in modeling Maine was also expected because of Maine's outlier behavior in ILI, having values considerably lower and out of pattern with other states.} The models in Fig \ref{states} included all 10 pages aggregated together. However, as indicated by the individual state results, this does not always lead to the best fit. Successful models often included a combination of select pages (such as FluView, Symptoms, and Treatment) but not an aggregation of all 10. Furthermore, aside from Texas, we did not have ILI data for the states outside of the typical flu season. Without this additional data, we are unable to determine how strongly the lower page views in the off-season correlate with off-season ILI. \\
\indent We then shifted the ILI data forward by one week. The regression analysis yielded 7 state/season combinations with $r^2$ values greater than 0.7 (see Table \ref{shiftstates}). The table also includes both the regular and normalized root mean squared errors
\begin{table}[H]
\begin{center}
\begin{tabular}{|c|c|c|c|c|}
\hline
\textbf{State} & \textbf{Season} & $\mathbf{r^2}$ & RMSE & NRMSE\\
\hline
Texas & 2012--2013 & 0.930 &0.667 & 0.067 \\
\hline
New Jersey & 2012--2013 & 0.832 & 0.767 & 0.117\\
\hline
New Mexico & 2015--2016 & 0.771 & 1.184 & 0.197\\
\hline
California & 2012--2013 & 0.746 & 0.797 & 0.129\\
\hline
Wisconsin & 2012--2013 & 0.727 & 0.626 & 0.153\\
\hline
North Carolina & 2015--2016 & 0.708 & 1.028 & 0.204\\
\hline
Missouri & 2012--2013 & 0.702 & 1.039 & 0.165\\
\hline
\end{tabular}
\end{center}
\caption{This table shows the states with models that had an $r^2 > 0.7$ when aggregating all 10 pages and shifting the ILI data forward by one week. The regular and normalized RMSEs are also displayed.}
\label{shiftstates}
\end{table}

Adding only the FluView, Symptoms, and Treatment pages,  we obtained $r^2 \geq 0.7$ for 6 state/season combinations. For the 2013--2014 season, the highest $r^2$ values were 0.612 for California and 0.568 for Wisconsin. While this is still less than desired, it is a vast improvement upon the $r^2$ values found from adding all 10 pages. For the 2014--2015 season, the highest $r^2$ was 0.575 for Missouri. Again, although the correlation appears to be weak, it is a stronger correlation than taking all 10 pages together. Using these same three pages and implementing a one-week shift, we obtained $r^2 \geq 0.7$ for 10 state/season combinations. For the 2014--2015 season, the highest $r^2$ value was 0.548 for Missouri. 

	\subsection{Model Failures}
	We generally found the models to be successful when considering pages most closely related to typical health-seeking behavior and when considering each flu season individually. When trying to model multiple influenza seasons together, we had a number of unsuccessful models. Considering all pages and national ILI data, the model combining the 2012-2013 and 2013-2014 influenza seasons had an $r^2$ of 0.061 and RMSE of 0.553. The combined 2013-2014 and 2014-2015 model had an $r^2$ of 0.241 and RMSE of 0.208. The combined 2014-2015 and 2015-2016 model had an $r^2$ of 0.251 and RMSE of 0.286. At the state level, combining all pages resulted in a number of unsuccessful models. For the 2013-2014 season, the Wisconsin model had an $r^2$ value of 0.187 and RMSE of 0.523. For the 2014-2015 season, the Missouri model had an $r^2$ value of 0.322 and RMSE of 1.845. \\
\indent We speculate that a number of factors could contribute to these negative results. While influenza is a seasonal disease, similar strains can span multiple years, affecting the susceptible populations in subsequent years. Our data stream may be biased toward individuals with more awareness of the CDC. Furthermore, individuals who search for influenza information in one season may not search for that information the next year. Finally, with the exception of Texas, we only have ILI data for the influenza season itself. Thus, while we do have Internet data for off-season influenza page views, we do not have corresponding ILI data.

\section{Conclusions}
Internet surveillance data has proven beneficial in predicting ILI incidence during flu seasons. However, our results show that the benefit of Internet data streams on informing disease is inconclusive. That is, our work shows that the CDC website traffic can be informative in some cases (e.g., national level) but not in others (e.g., state level). To determine the extent, we must return to our original research questions.

\vspace{.25cm}

\noindent
\textbf{Q1}: Given the successes of some of our models, we can conclude that CDC page view data can be used as an additional data source for monitoring disease incidence in some cases (for example, at the national level). The degree to which this data can be used appears to rely on the page selection and time frame. We obtained successful {nowcasts} when selecting pages related to topics most commonly used for online health queries (specific diseases and treatments) during the time span of a typical influenza season. Longer time spans and pages less associated with specific diseases and treatments led to less successful models. These results can assist others in selecting appropriate supplemental data sets for disease surveillance. \\

\vspace{.25cm}

\noindent
\textbf{Q2}: We obtained our most successful results using a one-week shift. Two-week shifts were successful in some cases but were overall less correlated than one-week shifts. Using no shift at all proved successful in some cases but not in others. We surmise that the shift required for the best fit depends upon the incubation period for the disease in question as well as the time period of reporting. The CDC Internet data are available daily; however, ILI data are available weekly, so we are limited in the types of shifts we can apply to the data sets. \\

\vspace{.25cm}

\indent We conclude that more studies on Internet data streams are needed to understand when and why Internet data works. Our methods are consistent with other feasibility studies and provide insight into conditions under which Internet data streams may inform influenza models. Future work should include rigorously testing the predictive power of the models by separating data into training and testing sets \cite{reid}.

\section*{Appendix A} \label{Appendix A}
\subsection*{CDC Pages}
This section contains the names and time frames for each page used. Some of the pages were renamed during the time period covered here.
\subsubsection*{Antivirals}
	\begin{table}[H]
	\begin{center}
	\begin{tabular}{|c|c|}
	\hline
	Dates & Page Name \\
	\hline
	12/30/12--10/20/13 & CDC--Seasonal Influenza (Flu)--Antivirals \\
	\hline
	10/20/13--8/10/14 & CDC--Antiviral Dosage $|$ Health Professionals $|$ \\
	& Seasonal Influenza (Flu) \\
	\hline
	8/24/14--5/28/16 & Antiviral Dosage $|$ Health Professionals $|$ Seasonal Influenza (Flu) \\
	\hline
	\end{tabular}
	\end{center}
	\end{table}

\subsubsection*{Flu Basics}
	\begin{table}[H]
	\begin{center}
	\begin{tabular}{|c|c|}
	\hline
	Dates & Page Name \\
	\hline
	12/30/12--10/13/13 & CDC--Seasonal Influenza (Flu)--Seasonal Influenza Q \& A: Flu Basics \\
	\hline
	10/13/13--6/8/14 & CDC--Seasonal Influenza: Flu Basics $|$ Seasonal Influenza (Flu) \\
	\hline
	6/8/14--7/27/14 & Seasonal Influenza: Flu Basics $|$ Seasonal Influenza (Flu) \\
	\hline
	7/27/14--8/31/14 & Seasonal Influenza: Flu Basics $|$ About (Flu) $|$ CDC \\
	\hline
	8/31/14--5/28/16 & Seasonal Influenza: Flu Basics $|$ Seasonal Influenza (Flu) $|$ CDC \\
	\hline
	\end{tabular}
	\end{center}
	\end{table}
	
\subsubsection*{FluView}
	\begin{table}[H]
	\begin{center}
	\begin{tabular}{|c|c|}
	\hline
	Dates & Page Name \\
	\hline
	12/30/12--10/26/14 & CDC--Seasonal Influenza (Flu)--FluView Interactive \\
	\hline
	10/26/14--5/28/16 & FluView Interactive $|$ Seasonal Influenza (Flu) $|$ CDC \\
	\hline
	\end{tabular}
	\end{center}
	\end{table}

\subsubsection*{High Risk Complications}
	\begin{table}[H]
	\begin{center}
	\begin{tabular}{|c|c|}
	\hline
	Dates & Page Name \\
	\hline
	12/30/12--10/13/13 & CDC--Seasonal Influenza (Flu)--People at High Risk\\
	 & of Developing Flu-Related Complications \\
	\hline
	10/13/13--6/8/14 & CDC--People at High Risk of Developing \\
	& Flu-Related Complications $|$ Seasonal Influenza (Flu)\\
	\hline
	6/8/14--7/27/14 & CDC--People at High Risk of Developing \\
	& Flu-Related Complications $|$ Seasonal Influenza (Flu)\\
	\hline
	8/17/14--5/28/16 & CDC--People at High Risk of Developing \\
	& Flu-Related Complications $|$ Seasonal Influenza (Flu) $|$ CDC\\
	\hline
	\end{tabular}
	\end{center}
	\end{table}
	
\subsubsection*{Key Facts}
	\begin{table}[H]
	\begin{center}
	\begin{tabular}{|c|c|}
	\hline
	Dates & Page Name \\
	\hline
	12/30/12--10/27/13 & CDC--Seasonal Influenza (Flu)--Key Facts\\
	& About Seasonal Flu Vaccine \\
	\hline
	10/27/13--7/13/14 & CDC--Key Facts About Seasonal Flu Vaccine $|$ \\
	& Seasonal Influenza (Flu)\\
	\hline
	7/13/14--7/27/14 & Key Facts About Seasonal Flu Vaccine $|$ \\
	& Seasonal Influenza (Flu) \\
	\hline
	7/27/14--5/28/16 & Key Facts About Seasonal Flu Vaccine $|$\\
	& Seasonal Influenza (Flu) $|$ CDC \\
	\hline
	\end{tabular}
	\end{center}
	\end{table}
	
\subsubsection*{Prevention}
	\begin{table}[H]
	\begin{center}
	\begin{tabular}{|c|c|}
	\hline
	Dates & Page Name \\
	\hline
	12/30/13--10/20/13 & CDC--Seasonal Influenza (Flu)--Prevention Strategies\\
	& for Seasonal Influenza in Healthcare Settings\\
	\hline
	10/27/13--6/1/14 & CDC--Prevention Strategies\\
	& for Seasonal Influenza in Healthcare Settings $|$ Healthcare\\
	& Professionals $|$ S...\\
	\hline
	6/8/14--5/28/16 & Prevention Strategies \\
	& for Seasonal Influenza in Healthcare Settings $|$ Health\\
	& Professionals $|$ Seasona...\\
	\hline
	\end{tabular}
	\end{center}
	\end{table}
	
\subsubsection*{Symptoms}
	\begin{table}[H]
	\begin{center}
	\begin{tabular}{|c|c|}
	\hline
	Dates & Page Name \\
	\hline
	12/30/12--10/13/13 & CDC--Seasonal Influenza (Flu)--Flu Symptoms \& Severity \\
	\hline
	10/20/13--6/8/14 & CDC--Flu Symptoms \& Severity $|$ Seasonal Influenza (Flu) \\
	\hline
	6/15/14--7/27/14 & Flu Symptoms \& Severity $|$ Seasonal Influenza (Flu) \\
	\hline
	7/27/14--8/31/14 & Flu Symptoms \& Severity $|$ About (Flu) $|$ CDC \\
	\hline
	9/7/14--5/28/16 & Flu Symptoms \& Severity $|$ Seasonal Influenza (Flu) $|$ CDC \\
	\hline
	\end{tabular}
	\end{center}
	\end{table}
	
\subsubsection*{Treating Influenza}
	\begin{table}[H]
	\begin{center}
	\begin{tabular}{|c|c|}
	\hline
	Dates & Page Name \\
	\hline
	12/30/12--2/28/16 & CDC--Seasonal Influenza (Flu)--Q \& A: Treating the Flu \\
	\hline
	\end{tabular}
	\end{center}
	\end{table}
	
\subsubsection*{Treatment}
	\begin{table}[H]
	\begin{center}
	\begin{tabular}{|c|c|}
	\hline
	Dates & Page Name \\
	\hline
	12/30/12--10/13/13 & CDC--Seasonal Influenza (Flu)--H3N2v Treatment \\
	\hline
	10/13/13--6/8/14 & CDC--Treatment--Antiviral Drugs $|$ Seasonal Influenza (Flu) \\
	\hline
	6/8/14--7/27/14 & Treatment--Antiviral Drugs $|$ Seasonal Influenza (Flu) \\
	\hline
	7/27/14--5/28/16 & Treatment--Antiviral Drugs $|$ Seasonal Influenza (Flu) $|$ CDC \\
	\hline
	\end{tabular}
	\end{center}
	\end{table}
	
\subsubsection*{Vaccine}
	\begin{table}[H]
	\begin{center}
	\begin{tabular}{|c|c|}
	\hline
	Dates & Page Name \\
	\hline
	12/20/12--9/7/14 & CDC--Seasonal Influenza (Flu)--Vaccination \\
	\hline
	8/31/14--5/28/16 & Cell-based Flu Vaccines $|$ Seasonal Influenza (Flu) $|$ CDC \\
	\hline
	\end{tabular}
	\end{center}
	\end{table}
\section*{Appendix B} \label{Appendix B}
\subsection*{State ILI Sources}
This section contains the sources used for the listed states' ILI percents.
\subsubsection*{California}
ILI data were obtained from the California Department of Public Health website at the following URL:
	\begin{itemize}
	\item http://www.cdph.ca.gov/data/statistics/Pages/CISPDataArchive.aspx
	\end{itemize}
	
\subsubsection*{Maine}
ILI data were obtained from the State of Maine Department of Health and Human Services website at the following URL:
	\begin{itemize}
	\item http://www.maine.gov/dhhs/mecdc/infectious-disease/epi/influenza/influenza-surveillance-archives.htm
	\end{itemize}
	
\subsubsection*{Missouri}
ILI data were obtained from the Missouri Department of Health \& Senior Services website at the following URL:
	\begin{itemize}
	\item http://health.mo.gov/living/healthcondiseases/communicable/influenza/reports.php
	\end{itemize}
\subsubsection*{New Jersey}
ILI data were obtained from the State of New Jersey Department of Health website at the following URLs:
	\begin{itemize}
	\item http://www.state.nj.us/health/flu/archives/shtml
	\item http://www.state.nj.us/health/flu/fluinfo.shtml
	\end{itemize}
	
\subsubsection*{New Mexico}
ILI data were obtained from the New Mexico Department of Health website at the following URL:
	\begin{itemize}
	\item https://nmhealth.org/about/erd/ideb/isp/data/
	\end{itemize}
	
\subsubsection*{North Carolina}
ILI data were obtained from the North Carolina Health and Human Services website at the following URLs:
	\begin{itemize}
	\item http://epi.publichealth.nc.gov/cd/flu/figures/flu1213.pdf
	\item http://epi.publichealth.nc.gov/cd/flu/figures/flu1314.pdf
	\item http://epi.publichealth.nc.gov/cd/flu/figures/flu1415.pdf
	\item http://flu.nc.gov/data/documents/flu1516.pdf
	\end{itemize}
	
\subsubsection*{Texas}
ILI data were obtained from the Texas Department of State Health Services website at the following URLs:
	\begin{itemize}
	\item www.dshs.texas.gov/idcu/disease/influenza/surveillance/2013/
	\item www.dshs.texas.gov/idcu/disease/influenza/surveillance/2014/
	\item www.dshs.texas.gov/idcu/disease/influenza/surveillance/2015/
	\end{itemize}
	
\subsubsection*{Wisconsin}
ILI data were provided by Thomas E. Haupt of Wisconsin Department of Health Services.
\section*{Appendix C} \label{Appendix C}
\subsection*{Census Divisions}
This section lists the states that comprise each of the U.S. census division.
\subsubsection*{East North Central}
	\begin{itemize}
	\item Illinois
	\item Indiana
	\item Michigan
	\item Ohio
	\item Wisconsin
	\end{itemize}
	
\subsubsection*{East South Central}
	\begin{itemize}
	\item Alabama
	\item Kentucky
	\item Mississippi
	\item Tennessee
	\end{itemize}
\subsubsection*{Middle Atlantic}
	\begin{itemize}
	\item New Jersey
	\item New York
	\item Pennsylvania
	\end{itemize}
	
\subsubsection*{Mountain}
	\begin{itemize}
	\item Arizona
	\item Colorado
	\item Idaho
	\item Montana
	\item New Mexico
	\item Nevada
	\item Utah
	\item Wyoming
	\end{itemize}
	
\subsubsection*{New England}
	\begin{itemize}
	\item Connecticut
	\item Maine
	\item Massachusetts
	\item New Hampshire
	\item Rhode Island
	\item Vermont
	\end{itemize}
	
\subsubsection*{Pacific}
	\begin{itemize}
	\item Alaska
	\item California
	\item Hawaii
	\item Oregon
	\item Washington
	\end{itemize}
	
\subsubsection*{South Atlantic}
	\begin{itemize}
	\item Delaware
	\item Florida
	\item Georgia
	\item Maryland
	\item North Carolina
	\item South Carolina
	\item Virginia
	\item West Virginia
	\end{itemize}
	
\subsubsection*{West North Central}
	\begin{itemize}
	\item Iowa
	\item Kansas
	\item Minnesota
	\item Missouri
	\item Nebraska
	\item North Dakota
	\item South Dakota
	\end{itemize}
	
\subsubsection*{West South Central}
	\begin{itemize}
	\item Arkansas
	\item Louisiana
	\item Oklahoma
	\item Texas
	\end{itemize}

\section*{Appendix D} \label{Appendix D}
\subsection*{Regional and National ILI}
The ILI data at both the census division and national levels are available in RegionalNationalILI.csv.

\section*{Appendix E} \label{Appendix E}
\subsection*{State ILI}
The ILI data for the states used in this study are available in StateILI.csv.

\end{document}